\documentclass[letterpaper, 10 pt, conference]{ieeeconf}
\IEEEoverridecommandlockouts
\usepackage{algorithmic}
\usepackage{algorithm}
\usepackage{amsmath,amssymb,amsfonts}

\usepackage{amsthm}

\usepackage{subcaption}
\usepackage[english]{babel}
\usepackage{graphicx}
\usepackage{textcomp}
\usepackage{xcolor}
\def\BibTeX{{\rm B\kern-.05em{\sc i\kern-.025em b}\kern-.08em
    T\kern-.1667em\lower.7ex\hbox{E}\kern-.125emX}}

\usepackage[font={sf,scriptsize},           
            caption=false]                  
           {subfig} 

\usepackage[capitalize]{cleveref}

\usepackage[
    style=ieee
]{biblatex}

\usepackage{subfiles}

\newtheorem{definition}{Definition}[section]

\newtheorem{lemma}{Lemma}[section]
           
\begin{document}

\title{Collision Avoidance Verification of \\Multiagent Systems with Learned Policies
}

\author{Zihao Dong$^1$, Shayegan Omidshafiei$^2$, Michael Everett$^1$
\thanks{$^1$Northeastern University, Boston, MA, USA. e-mail: \{\texttt{dong.zih, m.everett\}@northeastern.edu}. $^2$Work done while at Google.
\textbf{Code:} \protect\url{https://github.com/neu-autonomy/ReBAR}
}
}

\maketitle

\begin{abstract}
For many multiagent control problems, neural networks (NNs) have enabled promising new capabilities.
However, many of these systems lack formal guarantees (e.g.,  collision avoidance, robustness), which prevents leveraging these advances in safety-critical settings. While there is recent work on formal verification of NN-controlled systems, most existing techniques cannot handle scenarios with more than one agent. To address this research gap, this paper presents a backward reachability-based approach for verifying the collision avoidance properties of Multi-Agent Neural Feedback Loops (MA-NFLs). Given the dynamics models and trained control policies of each agent, the proposed algorithm computes \textit{relative backprojection sets} by (simultaneously) solving a series of Mixed Integer Linear Programs (MILPs) offline for each pair of agents. We account for state measurement uncertainties, making it well aligned with real-world scenarios. Using those results, the agents can quickly check for collision avoidance online by solving low-dimensional Linear Programs (LPs). We demonstrate the proposed algorithm can verify collision-free properties of a MA-NFL with agents trained to imitate a collision avoidance algorithm (Reciprocal Velocity Obstacles). We further demonstrate the computational scalability of the approach on systems with up to 10 agents.
\end{abstract}

\section{Introduction}

Verifying the properties of multiagent systems has been an important research area for several decades~\cite{pallottino2006probabilistic,doan2014verifying,kouvaros2019formal}.
Meanwhile, neural network (NN) control policies are becoming a key component of many state-of-the-art multiagent systems, such as for swarming \cite{tolstaya2020learning} and autonomous driving \cite{palanisamy2020multi}, yet the above verification algorithms cannot deal with these NNs.
Obtaining formal guarantees (e.g., for collision avoidance, robustness) for closed-loop systems with NN controllers, i.e., neural feedback loops (NFLs), remains challenging primarily due to the high dimensionality and nonlinearities of NNs.
Recent literature aims to provide these formal guarantees, typically by formulating reachability analysis problems and using convex relaxations to obtain tractable verification algorithms \cite{dutta2019reachability, huang2019reachnn, hu2020reach, everett2021reachability, rober2022backward, xu2020automatic}.
However, existing work focuses primarily on verification of scenarios with a single agent.

\begin{figure}[t]
    \centering
    \includegraphics[width=0.9\linewidth]{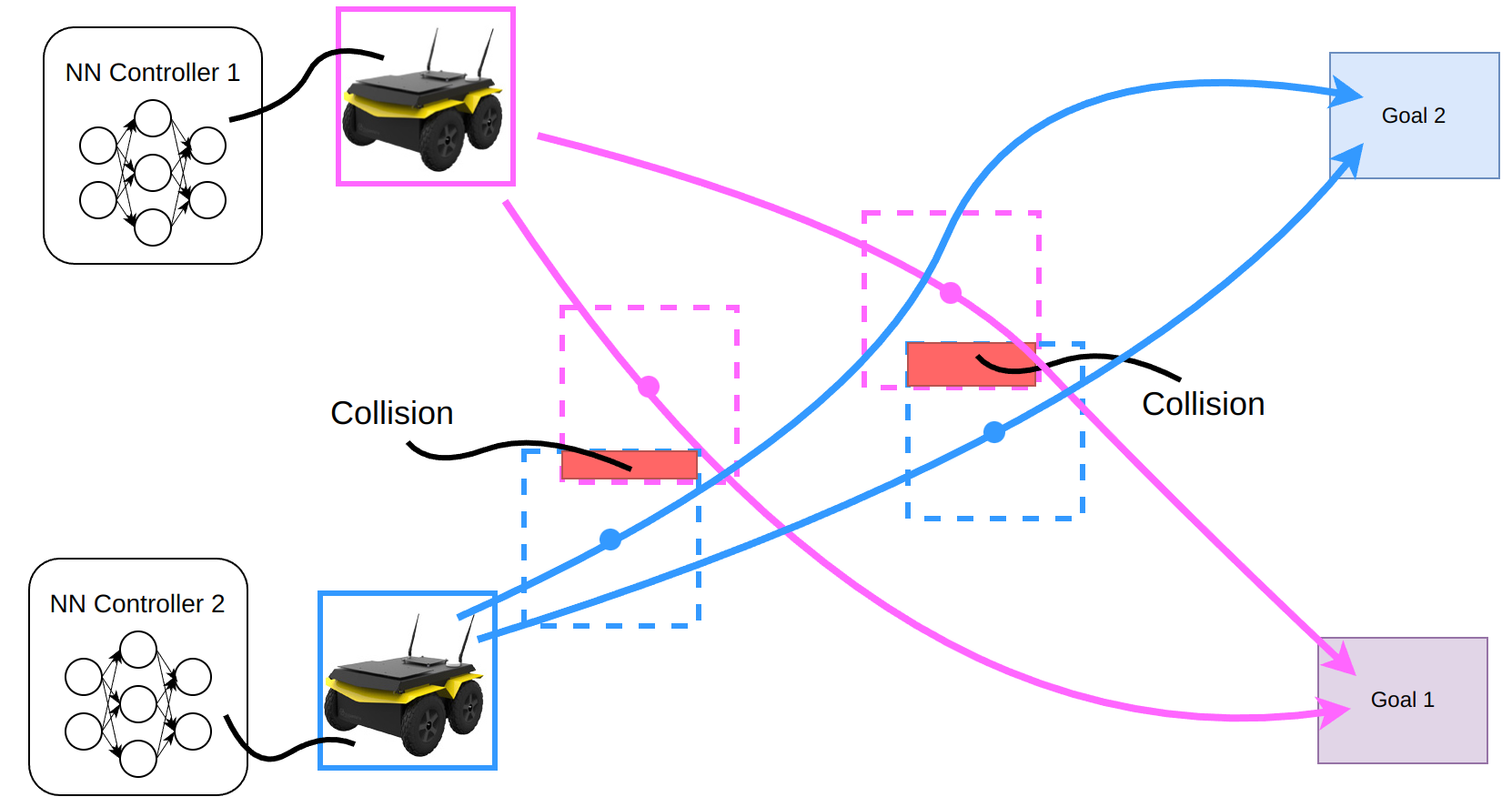}
    \caption{Complex interactions between agents present challenges in formal safety verification. This analysis is further complicated when the agents are using NN control policies.}
    \label{fig:cartoon}
    \vspace{-0.12in}
\end{figure}

\begin{figure}[t]
    \vspace{8pt}
    \captionsetup[subfigure]{justification=centering}
    \centering
    \begin{subfigure}[b]{0.5\linewidth}
    \includegraphics[width=0.87\linewidth]{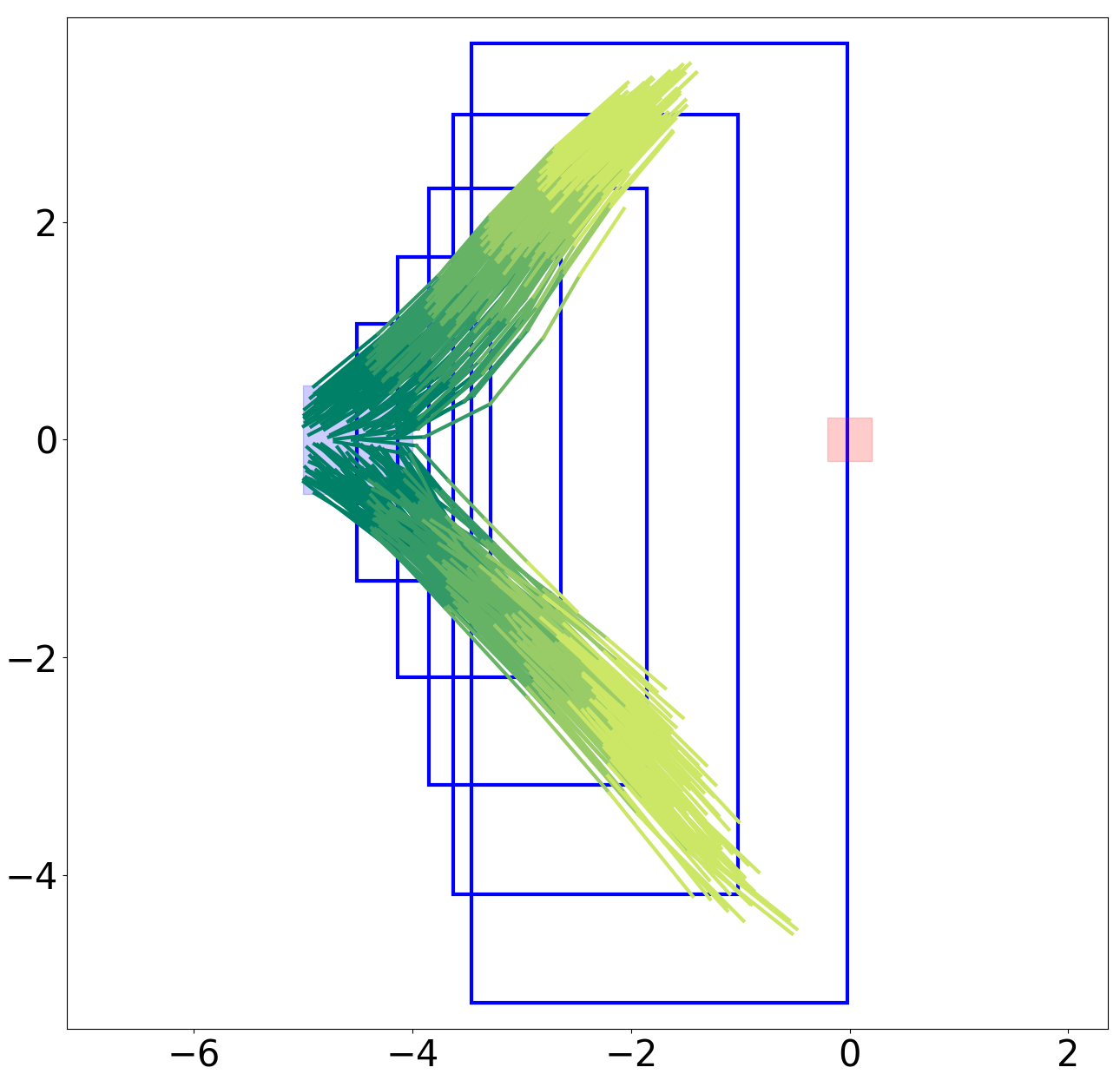}
    \caption{Forward Reachable Sets\\(Unable to verify collision avoidance)}
    \end{subfigure}%
    \begin{subfigure}[b]{0.45\linewidth}
    \includegraphics[width=\linewidth]{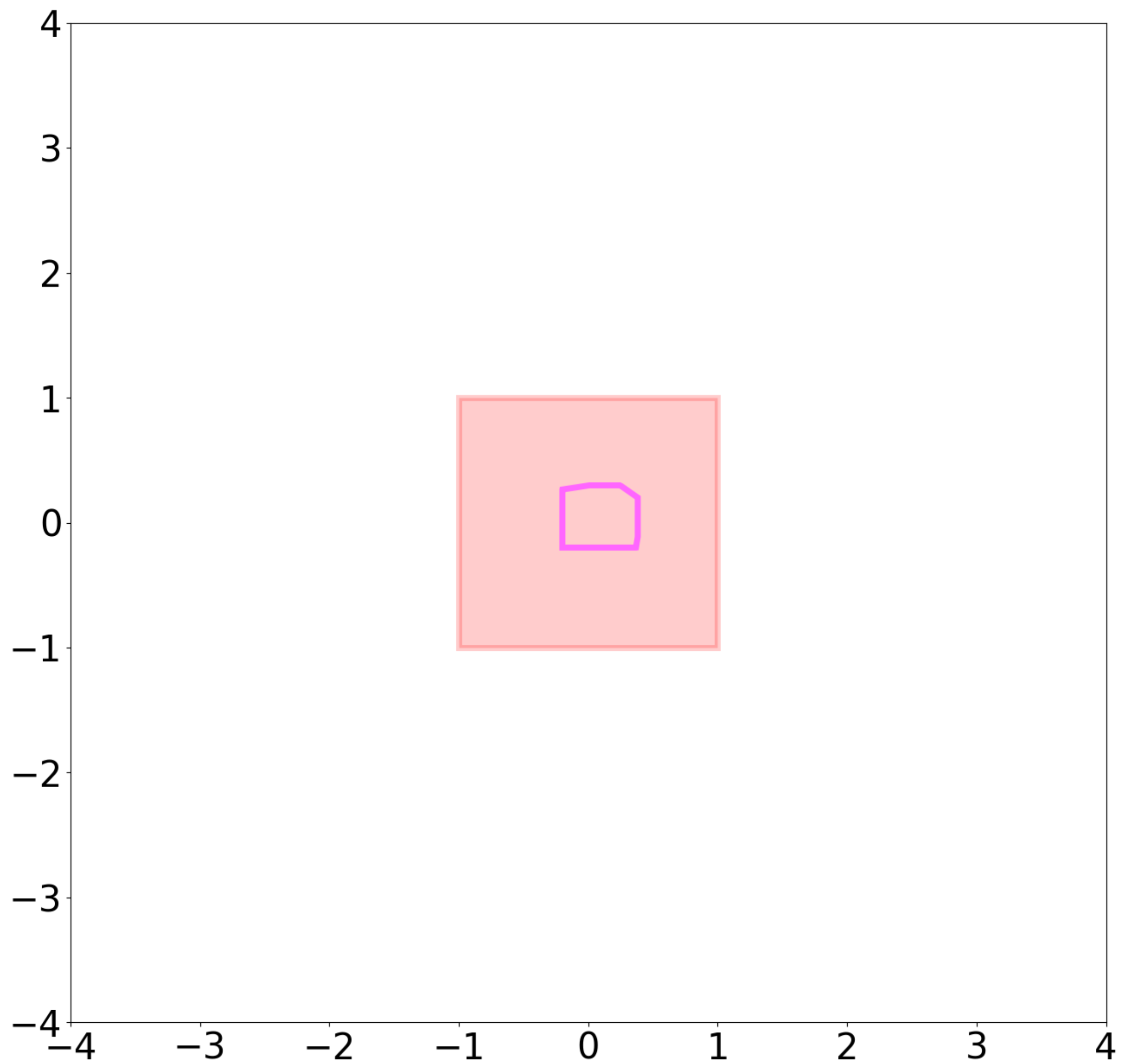}
    \caption[]{Relative Backprojection Sets\\(Able to verify collision avoidance)}
    \end{subfigure}
    \caption{(a): Forward reachable sets of agent 1 (blue) intersect with agent 2 (red). Unable to verify system safety despite no simulated rollouts (green lines) enters agent 2. (b): Result of ReBAR (magenta) is within agent 2, system is verified safe.}
    \label{fig:failure}
    \vspace{-0.15in}
\end{figure}

As shown in \cref{fig:cartoon}, extending these ideas to multiagent systems raises additional challenges, both in analyzing the complex multiagent interactions and in handling the increase in dimensionality.
Ref. \cite{yan2022strategy, yan2023partially} considers the verification of a two agent system with neuro-symbolic agents.
Ref. \cite{gates2023scalable} is the first approach for formally verifying multiagent NFLs (MA-NFLs).
In particular, \cite{gates2023scalable} extends Reach-SDP \cite{hu2020reach} to compute forward reachable sets over a single representation that contains all agents' controllers and dynamics.
While this approach provides meaningful bounds for several multiagent systems, our experiments suggest that systems with agents trained to avoid collisions are not well-suited for forward reachability-based techniques, especially when the given controllers are safe.
In \cref{fig:failure}, one agent uses the vector field policy from \cite{rober2022backward}, and the other agent is static at the origin (i.e., using a policy that always commands zero velocity).
Since the (convex) forward reachable sets (computed by merging both controllers as in \cite{gates2023scalable}, then solving mixed integer linear programs (MILPs)) contain both ways the agent could avoid the static agent, the reachable sets intersect with the obstacle, meaning the algorithm is unable to verify collision avoidance. 
Meanwhile, backward reachability analysis algorithms, such as BReach-LP~\cite{rober2022backward} (which can handle this scenario as a single-agent problem), prove the agents will not collide because although multiple disjoint paths might lead to the target set, these paths must remain within the target set. 
However, BReach-LP (and its recent extensions) do not handle more general multiagent systems (e.g., with multiple non-static agents).

Therefore, this paper aims to extend backward reachability analysis to MA-NFLs. 
Backward reachability-based algorithms typically compute backprojection sets of the avoid sets \cite{rober2022backward}. 
However, computing backprojection sets in the global frame is ineffective in the context of collision avoidance for MA-NFLs. 
Unlike in single agent setting where the avoid set is static, our avoid set is a moving region controlled by another neural network.
Computing its backprojection set in the global frame will be thus equivalent to computing the backprojection set of the whole state space.
In this work, we propose \textbf{Re}lative \textbf{BA}ckward \textbf{R}echability (ReBAR) and ReBAR-MA, computing \textit{relative backprojection sets} in a relative coordinate frame using MILPs.
This approach enables successful verification of the scenario in \cref{fig:failure}.
In particular, \cref{fig:failure}(b) shows that the relative backprojection set computed by ReBAR (magenta) lies within the collision set (red), meaning the agents can only collide if they start in a collision.
Furthermore, for systems that cannot be verified safe offline, we propose an efficient online safety check using LPs.
In summary, the contributions of this work include:
\begin{itemize}
    \item ReBAR: the first algorithm for formally verifying the collision avoidance properties of MA-NFLs, by computing relative backprojection sets with state uncertainty offline using MILPs and performing online safety checks online using low-dimensional LPs,
    \item ReBAR-MA: an extension of ReBAR to handle scenarios with arbitrary numbers of agents by performing parallelizable pair-wise verification, 
    \item an extension of them to provide safety guarantee over multiple timesteps in an iterative manner, and
    \item demonstrations of ReBAR and ReBAR-MA on relevant multiagent systems, including a 2-agent NFL trained to imitate the Reciprocal Velocity Obstacle (RVO) algorithm \cite{van2008reciprocal} and systems with up to 10 agents.
\end{itemize}

\section{Preliminaries}

\subsection{Multiagent System Dynamics}

For a system with $n$ agents, let the state of agent $i \in [n]$, where $[n]$ denotes the set $\{1, \dots, n\}$, be $\mathbf{x}^{(i)}_t \in \mathbb{R}^{n_x}$ and the joint system state, $\mathbf{X}_t \in \mathcal{X} \subseteq \mathbb{R}^{n * n_x}$, be
\begin{equation} \label{eq:concat}
    \mathbf{X}_t = [\mathbf{x}^{(1) \intercal}_t, \, \mathbf{x}^{(2) \intercal}_t, \, \cdots, \, \mathbf{x}^{(n) \intercal}_t]^\intercal.
\end{equation}
Throughout the paper,  superscripts $^{(i)}$ denote correspondence to agent $i$ (not derivatives).
Each agent's dynamics
can be described by the discrete-time, linear, time-invariant system,
\begin{align}\label{eq:dynamics}
    \begin{split}
        \mathbf{x}^{(i)}_{t+1} & = \mathbf{A}^{(i)} \mathbf{x}^{(i)}_t + \mathbf{B}^{(i)} \mathbf{u}^{(i)}_t + \mathbf{c}^{(i)} \\
        \mathbf{Y}_t & = \mathbf{C} \mathbf{X}_t + \mathbf{\mathcal{E}}
    \end{split}
\end{align}
where $\mathbf{A}^{(i)} \in \mathbb{R}^{n_x \times n_x}$, $\mathbf{B}^{(i)} \in \mathbb{R}^{n_x \times n_u}$, $\mathbf{C} \in \mathbb{R}^{n_x}$ are known system matrices, $\mathbf{c}^{(i)}$ is a known exogenous input, and $\mathbf{\mathcal{E}} = [\mathbf{\epsilon^{(0)}}^\intercal, \cdots, \mathbf{\epsilon^{(n)}}^\intercal]^\intercal \in \mathbb{R}^{n*n_x}$ is measurement uncertainty, resulting in uncertain state measurement $\mathbf{Y}_t \in \mathcal{X} \subseteq \mathbb{R}^{n*n_x}$, 
\begin{equation} \label{eq:concat_uncertain}
    \mathbf{Y}_t = [\mathbf{y}^{(1) \intercal}_t, \, \mathbf{y}^{(2) \intercal}_t, \, \cdots, \, \mathbf{y}^{(n) \intercal}_t]^\intercal.
\end{equation}

We assume uncertain state meansurement is within a $L_\infty$ ball centered at $\mathbf{x}^{(i)}_t$ with radius $\mathbf{\epsilon}^{(i)}$, i.e., $\mathbf{y}^{(i)}_t \in \mathcal{B}_\infty(\mathbf{x}^{(i)}_t, \mathbf{\epsilon}^{(i)}) $. Given control limits $\mathcal{U} \subseteq \mathbb{R}^{n_u}$ and state feedback policy $\pi^{(i)}: \mathcal{X} \to \mathcal{U}$, the control input for each agent $\mathbf{u}^{(i)}_t = \pi^{(i)}(\mathbf{Y}_t)$. The closed-loop dynamics for a single agent is denoted as:
\begin{equation}
    \mathbf{x}^{(i)}_{t+1} = f^{(i)} (\mathbf{X}_t; \mathcal{X}, \mathcal{U}, \mathbf{A}^{(i)}, \mathbf{B}^{(i)}, \mathbf{C}, \mathbf{c}^{(i)}, \pi^{(i)}, \epsilon^{(i)})
\end{equation}

Let $\mathbf{X}^{(i,j)}=[\mathbf{x}^{(i) \intercal}, \mathbf{x}^{(j) \intercal}]^\intercal$, the closed-loop dynamics for agents $i$ and $j$ is denoted
\begin{equation} \label{eq:f_dyn}
    \mathbf{X}^{(i,j)}_{t+1} = f^{(i,j)} (\mathbf{X}_t; f^{(i)}, f^{(j)}).
\end{equation}

\subsection{Neural Network Controller Architecture}\label{sec:preliminaries:nn_architecture}

This work assumes the control policies are feedforward NNs with $L$ hidden layers. Each layer $\ell \in [L+1]$ has $n_{\ell}$ neurons, weight $\mathbf{W}^{(i), \ell} \in \mathbb{R}^{n_{\ell+1} \times n_\ell}$ and bias $\mathbf{b}^{(i), \ell} \in \mathbb{R}^{n_{\ell+1}}$. The piecewise linear nonlinearity (e.g. ReLU) is denoted $\sigma^{(i), \ell}$. The control of agent $i$ is computed as:
\begin{align} \label{eq:dynamics}
    \begin{split}
       \mathbf{x}^{(i), 0} & = \mathbf{Y}_t \\
       \mathbf{z}^{(i), \ell} & = \mathbf{W}^{(i), \ell} \mathbf{x}^{(i), \ell} + \mathbf{b}^{(i), \ell}, \forall \ell \in [L] \\
       \mathbf{x}^{(i), \ell+1} & = \sigma^{(i), \ell}(\mathbf{z}^{(i), \ell}), \forall \ell \in [L-1] \\
       \pi^{(i)}(\mathbf{Y}_t) & = \mathbf{z}^{(i), L}.
    \end{split}
\end{align}
Control limit is incorporated as in \cite{gates2023scalable}. Accounting for general activations (tanh) or other architectures (convolution) using convex relaxations is left for future work \cite{xu2020automatic}.

\subsection{Collision Sets}

For two agents $i,j \in [n]$, $i \neq j$, assume their positions $(x^{(i)}, \, y^{(i)})$, $(x^{(j)}, \, y^{(j)})$ are elements of state vector $\mathbf{X}$. The \textit{collision set} for agents $i, j$ is defined as a convex polytope, \\
{\small
\begin{equation}\label{eq:col_linear}
    \mathcal{C}^{(i,j)} \triangleq \{\mathbf{X} \text{ } | \text{ } \mathbf{H} \mathbf{X}^{(i,j)} \leq \mathbf{d} \}, 
\end{equation}
}
where $\mathbf{H} \in \mathbb{R}^{n_f \times 2n_x}$ and $\mathbf{b} \in \mathbb{R}^{n_f}$, and $n_f$ denotes the number of facets bounding this convex polytope.

\subsection{Backreachable \& Backprojection Set}

We extend the backprojection set definition in \cite{rober2022backward} to multiagent systems with state uncertainty, to handle avoid sets parameterized by coordinates of moving agents (instead of static obstacles). For agents $i,j \in [n]$, the \textit{backreachable set} contains all states such that $\exists \mathbf{u}^{(i)}_t, \mathbf{u}^{(j)}_t \in \mathcal{U}$ that drive the system into collision set on the next timestep.

{\footnotesize
    \begin{align} \label{eq:breach_set}
        \mathcal{R}_{-1}(\mathcal{C}^{(i,j)}) \triangleq \left\{ \mathbf{X}_t \middle\vert \begin{array}{l}
             \mathbf{x}^{i}_{t+1} =\mathbf{A}^{(i)} \mathbf{x}^{(i)}_t + \mathbf{B}^{(i)} \mathbf{u}^{(i)}_t + \mathbf{c}^{(i)} \\
             \mathbf{x}^{j}_{t+1} =\mathbf{A}^{(j)} \mathbf{x}^{(j)}_t + \mathbf{B}^{(j)} \mathbf{u}^{(j)}_t + \mathbf{c}^{(j)} \\
             \mathbf{u}^{(i)}_t, \mathbf{u}^{(j)}_t \in \mathcal{U} \\
             \mathbf{X}^{(i,j)}_{t+1} \in \mathcal{C}^{(i,j)}
        \end{array} 
        \right\}.
    \end{align}
}
And similarly, the \textit{backprojection set} contains all states such that the policies $\pi^{(i)}, \pi^{(j)}$ will drive the system into the collision set on the next timestep,

{\footnotesize
    \begin{align} \label{eq:bproj_set}
        \begin{split}
            \mathcal{P}_{-1}(\mathcal{C}^{(i,j)}) \triangleq \left\{ \mathbf{X}_t \text{ } \middle\vert \begin{array}{l}
                 \mathbf{X}^{(i,j)}_{t+1} = f^{(i,j)} (\mathbf{X}_t; f^{(i)}, f^{(j)}) \\
                 \mathbf{X}^{(i,j)}_{t+1} \in \mathcal{C}^{(i,j)}
            \end{array}
            \right\}.
        \end{split}
    \end{align}
}

\section{Approach}

\subsection{Relative Backreachable \& Backprojection Sets}

To compute safety certificates offline and use them online for safety verification, we compute backprojection sets of the collision set using the agents' relative coordinates. Let $\mathbf{p}^{(i)}_t = (x^{(i)}_t, y^{(i)}_t)$ denote the position of agent i at timestep t, and let $\mathbf{p}^{(j \to i)}_t = (x^{(j)}_t-x^{(i)}_t, y^{(j)}_t-y^{(i)}_t)$ denote the relative position of agent $j$ w.r.t. agent $i$. The collision set can be expressed using the relative coordinates as:
\begin{equation}
    \mathcal{C}^{(i,j)} \triangleq \{\mathbf{p}^{(j \to i)} \text{ } | \text{ } \mathbf{H} \mathbf{X}^{(i,j)} \leq \mathbf{d} \}.
    \label{eq:relative_col}
\end{equation}
\cref{eq:relative_col} is the failure set we consider in this work. The \textit{relative backreachable \& backprojection set} are defined as:

{\footnotesize
    \begin{align}
        \mathcal{R}_{-1}(\mathcal{C}^{(i,j)}) & \triangleq \left\{ \mathbf{p}^{(j \to i)}_t \middle\vert \begin{array}{l}
             \mathbf{x}^{i}_{t+1} =\mathbf{A}^{(i)} \mathbf{x}^{(i)}_t + \mathbf{B}^{(i)} \mathbf{u}^{(i)}_t + \mathbf{c}^{(i)} \\
             \mathbf{x}^{j}_{t+1} =\mathbf{A}^{(j)} \mathbf{x}^{(j)}_t + \mathbf{B}^{(j)} \mathbf{u}^{(j)}_t + \mathbf{c}^{(j)} \\
             \mathbf{u}^{(i)}_t, \mathbf{u}^{(j)}_t \in \mathcal{U} \\
             \mathbf{p}^{(j \to i)}_{t+1} \in \mathcal{C}^{(i,j)}
        \end{array} \right\} \\
        \mathcal{P}_{-1}(\mathcal{C}^{(i,j)}) & \triangleq \left\{ \mathbf{p}^{(j \to i)}_t \middle\vert \begin{array}{l}
             \mathbf{X}^{(i,j)}_{t+1} = f^{(i,j)} (\mathbf{X}_t; f^{(i)}, f^{(j)}) \\
             \mathbf{p}^{(j \to i)}_{t+1} \in \mathcal{C}^{(i,j)}
        \end{array} \right\}
    \end{align}
}

\begin{figure}[t] 
    \centering
    \includegraphics[width=0.65\columnwidth]{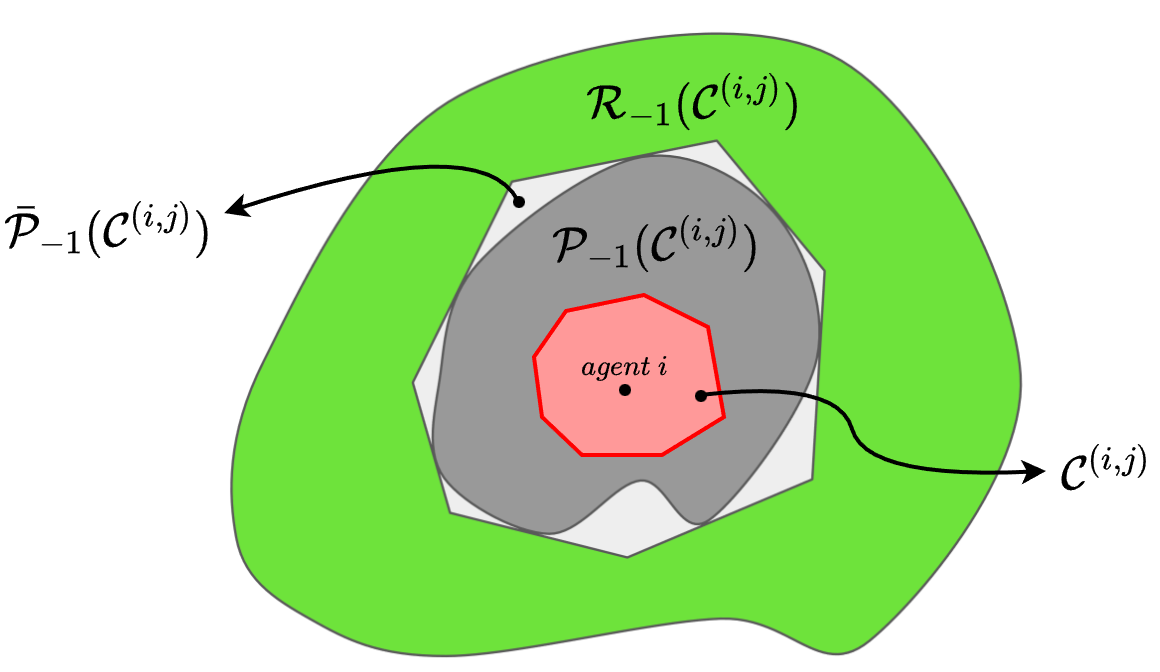}
    \caption{$\mathcal{C}^{(i,j)}$ (red) is a convex polytope around agent $i$; $\bar{\mathcal{P}}_{-1}(\mathcal{C}^{(i,j)})$ (light grey) is the tightest convex polytope of $\mathcal{P}_{-1}(\mathcal{C}^{(i,j)})$ (grey) using the given facets; $\mathcal{R}_{-1}(\mathcal{C}^{(i,j)})$ (green) contains all states that $\exists u \in \mathcal{U}$ s.t. agent $i,j$ collides}
    \label{fig:relative_bpset}
    \vspace{-0.1in}
\end{figure}

\cref{fig:relative_bpset} illustrates the relationship between collision set, relative backprojection set, relative backprojection set over approximation (RBPOA), and relative backreachable set. 
Unlike prior work~\cite{rober2022backward}, these sets are defined relative to one agent to be more suitable for analyzing collision avoidance of MA-NFLs.

\subsection{Two Agent Verification}

Because the relative backprojection sets can be non-convex and involve potentially high-dimensional NNs, computing these sets exactly is computationally intractable. Instead, we will compute a convex over approximation of these sets using finitely many facets $\mathbf{a} \in \mathbb{R}^2$. We begin by considering a system with two agents. For each facet, the following optimization problem finds a half-space containing the relative backprojection set:
\begin{align}\label{eq:two_agent}
    \begin{split}
    \underset{\mathbf{x}_t}{\text{min}} \quad & \mathbf{a}^\intercal \mathbf{p}^{(j \to i)}_t \\
        \text{s.t.} \quad & \mathbf{X}_t \in \mathcal{X} \\
             & \mathbf{X}^{(i,j)}_{t+1} = f^{(i,j)} (\mathbf{X}_t; f^{(i)}, f^{(j)}) \\
             & \mathbf{p}^{(j \to i)}_{t+1} \in \mathcal{C}^{(i,j)},
    \end{split}
\end{align}
and the intersection of the resulting half-spaces is a convex over-approximation of the relative backprojection set. Because we have assumed the neural network controllers have piecewise linear activation functions, the optimization problem \cref{eq:two_agent} is a MILP, with the same number of binary variables as the number of (uncertain) ReLU neurons in the neural network controller. The solutions to the MILPs provide the tightest convex over approximation using the given facets (subjected to numerical tolerance). In \cref{eq:two_agent}, t is an arbitrary timestep, meaning that the resulting convex polytope $\bar{\mathcal{P}}_{-1}(\mathcal{C}^{(i,j)})$ is an over approximation of the relative backprojection set at any arbitrary timestep $t \geq 0$. By utilizing the relative coordinate frame, we make implicit where the collision occurs globally, allowing to move the expensive RBPOA computation offline. \cref{alg:ReBAR} summarizes the approach for verifying a 2-agent NFL. Computing the RBPOA requires solving $n_f$ MILPs in total.

\begin{algorithm}[H]
    \caption{ReBAR}
    \begin{algorithmic}[1] \label{alg:ReBAR}
        \renewcommand{\algorithmicrequire}{\textbf{Input:}}
        \renewcommand{\algorithmicensure}{\textbf{Output:}}
        \REQUIRE $\text{collision set } \mathcal{C}^{(i,j)},$ \text{closed-loop dynamics function } $f^{(i,j)}$, \text{number of facets } $n_f$ \\ 
        \ENSURE $\text{RBPOA } \bar{\mathcal{P}}_{-1}(\mathcal{C}^{(i,j)})$ \\
        \STATE $\bar{\mathcal{P}}_{-1}(\mathcal{C}^{(i,j)}) \leftarrow \mathbb{R}^2$
        \FOR{ $\mathbf{a} \in \text{getFacets(}n_f\text{)}$}
            \STATE $b \leftarrow \text{solveForFacet(} \mathbf{a}, f^{(i,j)} \text{)} \quad (\cref{eq:two_agent})$
            \STATE $\mathcal{A} \leftarrow \{ { \mathbf{p}^{(j \to i)}} \text{ } | \text{ } \mathbf{a}^\intercal \mathbf{p}^{(j \to i)} \geq b \}$
            \STATE $\bar{\mathcal{P}}_{-1}(\mathcal{C}^{(i,j)}) \leftarrow \bar{\mathcal{P}}_{-1}(\mathcal{C}^{(i,j)}) \cap \mathcal{A} $
        \ENDFOR
        \STATE $\text{return }\bar{\mathcal{P}}_{-1}(\mathcal{C}^{(i,j)})$
    \end{algorithmic}
\end{algorithm}

Similar to \cite{rober2022backward}, the 2-agent system is verified safe, i.e. the agents will not collide at any future timestep if they do not collide at $t=0$, if the resulting RBPOA is a subset of the collision set, as formalized next.

\begin{definition}[Verified Safe] \label{dfn:verified_safe}
    A system with 2 agents $i,j$ is verified safe w.r.t. $\mathcal{C}^{(i,j)}$ if $\mathbf{p}^{(j \to i)}_t \notin \mathcal{C}^{(i,j)} \implies  \forall \tau \geq 0, \mathbf{p}^{(j \to i)}_{t+\tau} \notin \mathcal{C}^{(i,j)}$.
\end{definition}

\begin{lemma}\label{thm:verified_safe}
     A system with 2 agents $i,j$ with closed-loop dynamics function $f^{(i,j)}$ and collision set $\mathcal{C}^{(i,j)}$ is verified safe if $\bar{\mathcal{P}}_{-1}(\mathcal{C}^{(i,j)}) \subseteq \mathcal{C}^{(i,j)}$.
\end{lemma}

\begin{proof}
    Consider a 2-agent system satisfying $\bar{\mathcal{P}}_{-1}(\mathcal{C}^{(i,j)}) \subseteq \mathcal{C}^{(i,j)}$. Let $\mathbf{X}_0 \in \mathcal{X}, \mathbf{p}^{(j \to i)}_0 \notin \mathcal{C}^{(i,j)}$ be the starting state of the system, then $\mathbf{p}^{(j \to i)}_0 \notin \bar{\mathcal{P}}_{-1}(\mathcal{C}^{(i,j)})$, and thus $\mathbf{p}^{(j \to i)}_1 \notin \mathcal{C}^{(i,j)}$. At timestep $k \geq 0$, assume $\mathbf{X}_k \in \mathcal{X}$ and $\mathbf{p}^{(j \to i)}_k \notin \mathcal{C}^{(i,j)}$, thus $\mathbf{p}^{(j \to i)}_k \notin \bar{\mathcal{P}}_{-1}(\mathcal{C}^{(i,j)})$, and $\mathbf{p}^{(j \to i)}_{k+1} \notin \mathcal{C}^{(i,j)}$. By induction, this analysis can be extended to any arbitrary timestep $t \geq 0$, hence a two agent system is safe (from collision) if $\bar{\mathcal{P}}_{-1}(\mathcal{C}^{(i,j)}) \subseteq \mathcal{C}^{(i,j)}$.
\end{proof}

When a 2-agent NFL is verified safe by \cref{thm:verified_safe}, the agents will not collide as long as they start from a non-colliding state, even if they have noisy state measurement. However, a system that is not verified safe by \cref{thm:verified_safe} may still provide collision-free behavior in a subset of $\mathcal{X}$.
In practice, if only a noisy state estimate is available (e.g., uncertainty bounds over $\mathbf{x}_t$), it is not immediately obvious whether $ \mathbf{p}^{(j \to i)}$ lies in the RBPOA. Let the convex polytope $\mathbf{A}^{(i,j)}_t \mathbf{p}^{(j \to i)} \leq \mathbf{b}^{(i,j)}_t$ denote the uncertain state, we solve the following LP:
\begin{align}\label{eq:online}
    \begin{split}
        \text{min} & \quad 0 \\
        s.t. & \quad \mathbf{X}_t \in \mathcal{X}, \\
             & \quad \mathbf{A}^{(i,j)}_t \mathbf{p}^{(j \to i)} \leq \mathbf{b}^{(i,j)}_t \\
             & \quad { \mathbf{p}^{(j \to i)}} \in \bar{\mathcal{P}}_{-1}(\mathcal{C}^{(i,j)}_{t})
    \end{split}
\end{align}

If \cref{eq:online} is infeasible, then, even with measurement uncertainty, the system is guaranteed collision-free on the next timestep because ReBAR computes an over approximation of the true unsafe region. Note \cref{eq:online} is a LP that does not involve NN controllers, as thus can be solved efficiently, enabling our method to provide a safety guarantee online.

ReBAR can be extended to provide safety guarantees over an extended time horizon $\tau$ by computing RBPOAs at multiple timesteps $\bar{\mathcal{P}}_{-\tau:0}(\mathcal{C}^{(i,j)})$. We initialize the zeroth RBPOA as the collision set and step backward in time to compute RBPOA recursively using \cref{alg:ReBAR} and the previous timestep RBPOA as collision set. Computing $\bar{\mathcal{P}}_{-k}(\mathcal{C}^{(i,j)})$ using \cref{alg:ReBAR} only involves a change of variable (replacing $\mathcal{C}^{(i,j)}$ with $\bar{\mathcal{P}}_{-k+1}(\mathcal{C}^{(i,j)})$), so every step backward in time incurs similar time complexity.

\subsection{Scaling to More Agents}

For systems with more than 2 agents, considering all agents in one optimization problem (as in \cite{gates2023scalable}) becomes computationally intractable because the dimensionality of the problem will be prohibitive for verification as the number of agents increases. Instead, we propose to split the task into $\mathcal{O}(n^2)$ pair-wise verification sub-problems, each of which is the same as \cref{eq:two_agent} and can be solved using \cref{alg:ReBAR} in similar runtime as the two agent verification problem. Furthermore, the sub-problems only involve controllers of agent $i$ and $j$, and thus can be solved simultaneously across available resources, allowing our approach to scale to systems with larger numbers of agents. The approach is summarized in \cref{alg:MA}. Note this algorithm can be extended to $n$-step case similar as \cref{alg:ReBAR}. When every pair of agents is verified safe by \cref{thm:verified_safe}, then the system is verified safe, i.e., no agents will collide if they start from non-colliding states, as we formalize next.

\begin{definition}[Multi-Agent Verified Safe] \label{dfn:MA_verified_safe}
    A system with $n$ agents is verified safe if $\forall i,j \in [n], i \neq j, \mathbf{p}^{(j \to i)}_t \notin \mathcal{C}^{(i,j)} \implies  \forall \tau \geq 0, \mathbf{p}^{(j \to i)}_{t+\tau} \notin \mathcal{C}^{(i,j)}$
\end{definition}
\begin{lemma} \label{thm:MA_verified_safe}
     For a system with $n$ agents, closed-loop dynamics functions $f^{(0,1)}, f^{(0,2)}, \cdots, f^{(n-1,n)}$, and collision sets $\mathcal{C}^{(0,1)}, \mathcal{C}^{(0,2)}, \cdots, \mathcal{C}^{(n-1,n)}$. If $\bar{\mathcal{P}}_{-1}(\mathcal{C}^{(i,j)}) \subseteq \mathcal{C}^{(i,j)} \text{ } \forall i,j \in [n]$, the multi-agent system is verified safe.
\end{lemma}
\begin{proof}
    For a $n$-agent system that is pair-wise verified safe by \cref{thm:verified_safe}, let $\mathbf{X}_0 \in \mathcal{X}, \mathbf{p}^{(j \to i)}_0 \notin \mathcal{C}^{(i,j)} \text{ } \forall i,j \in [n], i \neq j,$ be the starting state. Assume $\forall i,j \in [n], i \neq j$, at timestep $k \geq 0$, $\mathbf{X}_k \in \mathcal{X}$ and $\mathbf{p}^{(j \to i)}_k \notin \mathcal{C}^{(i,j)}$. Because $\forall i,j \in [n], i \neq j, \text{ }$the relative backprojection set over approximation between agent $i,j$ is a subset of their collision set, i.e. $ \bar{\mathcal{P}}_{-1}(\mathcal{C}^{(i,j)}) \subseteq \mathcal{C}^{(i,j)} $, we must have $\forall i,j \in [n], i \neq j, \text{ } \mathbf{p}^{(j \to i)}_k \notin \bar{\mathcal{P}}_{-1}(\mathcal{C}^{(i,j)})$, and $\mathbf{p}^{(j \to i)}_{k+1} \notin \mathcal{C}^{(i,j)}$, i.e., no collision at timestep $k+1$. Proof completed by induction.
\end{proof}

\begin{algorithm}[H]
    \caption{ReBAR-MA}
    \begin{algorithmic}[1] \label{alg:MA}
        \renewcommand{\algorithmicrequire}{\textbf{Input:}}
        \renewcommand{\algorithmicensure}{\textbf{Output:}}
        \REQUIRE $\text{collision sets } \mathcal{C}^{(1,2)}, \cdots, \mathcal{C}^{(n-1,n)}$ \\ 
            \quad $ \text{closed-loop dynamic functions } f^{(1,2)}, \cdots, f^{(n-1,n)}$\\
            \quad $ \text{number of facets } n_f $ \\ 
        \ENSURE $\text{RBPOAs } \{ \bar{\mathcal{P}}_{-1}(\mathcal{C}^{(1,2)}), \dots, \bar{\mathcal{P}}_{-1}(\mathcal{C}^{(n-1, n)})\}$ \\
        \STATE $\bar{\mathcal{P}} = \{\}$ \\
        \# In Parallel \\
        \FOR{$i,j \in [n]$} 
            \STATE $\bar{\mathcal{P}}_{-1}(\mathcal{C}^{(i,j)}) \leftarrow \text{ReBAR}( \cdot )$ 
            \STATE $\bar{\mathcal{P}}.append(\bar{\mathcal{P}}_{-1}(\mathcal{C}^{(i,j)}))$
        \ENDFOR
        \STATE $\text{return } \bar{\mathcal{P}} $
    \end{algorithmic}
\end{algorithm}
\vspace{-0.1in}

In case we cannot verify the safety of the system, we can use the RBPOAs online to verify safety of states similar as the two agent case. If \cref{eq:online} is infeasible for all pairs of agents, then the state of the system is safe and the proof is similar to our proof of \cref{thm:MA_verified_safe}.

\section{Experiment}

All results are obstained using the optimization solver Gurobi \cite{gurobi} on a PC running Ubuntu 22.04 with a Intel i9-13900K CPU. ReBAR verifies a pair of agents mimicing RVO \cite{van2008reciprocal} offline in 11s and run safety check online in 1.4ms on average. We computed RBPOAs up to 5 steps, and visualize them with the RBPUAs we generated by exhaustive sampling. For systems with up to 10 agents trained to mimic the vector field in \cite{rober2022backward} with [20,20] hidden neurons, ReBAR-MA verifies each pair of agents in 200s to 320s, and the online check takes less than 1.6ms for each pair of agents.

\subsection{Two Agent Verification}

\subsection{Runtime and Scalability}

\section{Conclusion}

This paper presented ReBAR and ReBAR-MA for verifying the collision avoidance safety of Multi-Agent Neural Feedback Loops (MA-NFLs) by computing the relative backprojection set over approximation with state uncertainty offline, and using the result to provide a real-time safety guarantee. We demonstrate them by verifying agents trained mimicing RVO \cite{van2008reciprocal}, and system with up to 10 agents.
Future work will consider algorithmic extensions to handle general activation functions and dynamics through linear relaxation and verification in decentralized observation space.


\printbibliography

\end{document}